\documentclass[twocolumn,prl,aps,floatfix,showpacs]{revtex4}
\usepackage{amssymb}
\usepackage{epsfig}

\usepackage{amsmath}
\usepackage{graphicx}
\usepackage{dcolumn}
\usepackage{bm}

\begin{document}

\preprint{APS/123-QED}
\title{Optical Control of Topological Quantum Transport in Semiconductors}
\author{Wang Yao}
\thanks{Electronic address: wangyao@physics.utexas.edu}
\author{A. H. MacDonald}
\author{Qian Niu}
\affiliation{Department of Physics, The University of Texas, Austin,
Texas 78712}
\date{\today}

\begin{abstract}
Intense coherent laser radiation red-detuned from absorption edge
can reactively activate sizable Hall type charge and spin transport
in n-doped paramagnetic semiconductors as a consequence of $k$-space
Berry curvature transferred from valence band to photon-dressed
conduction band. In the presence of disorder, the optically induced
Hall conductance can change sign with laser intensity.
\end{abstract}

 \maketitle


Modification of material properties by external fields is of
fundamental interests in condensed matter physics. Well known
examples include optical Kerr nonlinearity, magneto-resistance, and
the Hall effect. These phenomena enable externally controllable
devices with importance in technologies. Semiconductors offer an
ideal platform to explore control possibilities due to rich
behaviors under magnetic, electric, and optical perturbations. In
particular, with advances in laser technology, optically activated
novel properties in semiconductors can now be explored with intense
coherent lasers~\cite{ORKKY} and are motivated by the need for
ultra-fast switchable devices.

In this letter, we predict coherently photoinduced Hall type charge
and spin transport in n-doped paramagnetic semiconductors in which
conduction band spin-orbit coupling can be absent. In
semiconductors, an optical field couples the lowest $s$-like
conduction band and the top $p$-like valance bands with a sizable
interband matrix element. When the light is far red-detuned in
frequency from the absorption edge, the influence on the electronic
system is reactive, rather than dissipative as in most previous
photoinduced transport
schemes~\cite{Spingalvanic_Ganichev,Photonspincurrent_Smirl,Photoinduced_Hall_Zhang}.
The off-resonance coherent laser beam renormalizes the band
structure~\cite{OpticalStark}, in a fashion analogous to the
dynamical Stark effect in atomic physics. $k$-space Berry curvatures
which originate from strong spin-orbit coupling in the valance bands
are redistributed in the photon-dressed bands and the occupied
electronic states acquires non-trivial topological
properties~\cite{QHE_TKNN,Chang_BerryPhase}. Circularly polarized
light breaks time reversal symmetry and leads to an anomalous Hall
effect
(AHE)~\cite{Karplus_AHE,Smit_skew1,Berger_Sidejump,AHE_FMSM_Jungwirth,AHE_Fang,AHE_Niu,AHE_ins_ext_Nagaosa}.
With linearly polarized light, time reversal symmetry forbids a net
charge Hall current, but a DC electric field can drive a pseudo-spin
current in the perpendicular direction, i.e. bidirectional Hall flow
of the dressed electrons similar to the spin Hall effect
(SHE)~\cite{SHE_Murakami,SHE_Awschalom,SHE_Wunderlich}. The
population proportion of the two species of electrons is driven into
imbalance at the edges of the light irradiation area, which results
in long-lived pure spin accumulations of conduction band electrons
when the light is turned off adiabatically.

The AHE has been one of the most fundamental yet controversial
issues in condensed matter physics. Despite its extensive use for
sample characterization in ferromagnetic systems, there have long
been debates if the AHE originates entirely from spin-dependent
scattering at impurities, or has important topological contribution
from $k$-space Berry
curvature~\cite{Karplus_AHE,Smit_skew1,Berger_Sidejump,AHE_FMSM_Jungwirth,AHE_Fang,AHE_Niu,AHE_ins_ext_Nagaosa}.
The SHE offers a promising avenue towards semiconductor spintronics
based on pure spin transport, yet the definition of spin current
remains a controversial issue in strong spin-orbit coupled
systems~\cite{Spincurrent_Niu}. Not coincidentally, it has been a
challenge to relate the spin current to the experimentally
observable spin accumulations at sample edges, since it has been
realized that the latter can also be induced locally and is
dependent on boundary conditions.

As an attempt to contribute to these conceptual debates that are
critical for making the connection between experimental and
theoretical studies, we choose to illustrate our idea by applying it
to an n-doped quantum well (QW) of III-V compounds where structure
inversion asymmetry (SIA) induced spin-orbit coupling is
absent~\cite{Rashba_1984}. We predict that the AHE and SHE can be
activated/deactivated in this paramagnetic system by switching
on/off the light field only, which can be a clear demonstration of
the topological contributions. This system has the unique feature
that it consists of two decoupled subspaces related by a
time-reversal operation~\cite{twospace}, which unambiguously leads
to the definition of the pseudo-spin current as the only source of
the spin accumulations with out-of-plane polarization at the light
irradiation edges.

We first explain the band structure renormalization by light for
noninteracting electrons and then consider the effect of
interactions. The explicit time dependence in the coupling with a
coherent laser field of frequency $\omega_0$ is removed by a
canonical transform $U(t)=\rm{exp} (-i \omega_0 t \sum_{\bf k}
a^{\dagger}_{c,{\bf k}} a_{c,{\bf k}})$, so that in this rotating
frame (RF), the Hamiltonian reads~\cite{OpticalStark}
\begin{eqnarray}
H_{\rm RF} &=& \sum_{{\bf k}} (\varepsilon_c({\bf k})-
\omega_0)a^{\dagger}_{c,{\bf k}} a_{c,{\bf k}} +\sum_{{\bf k},j}
\varepsilon_{v_j}({\bf k})
a^{\dagger}_{{v_j},{\bf k}} a_{{v_j},{\bf k}} \nonumber \\
&+&\sum_{\sigma} \sum_{{\bf k}, j} \Omega_{\sigma,j}({\bf k})
a^{\dagger}_{c,{\bf k}} a_{v_j,{\bf k}} + c.c. ~~,
\label{H_nocoulomb}
\end{eqnarray}
where $a^{\dagger}_{c,{\bf k}}$ and $a^{\dagger}_{v_j,{\bf k}}$ are
the particle creation operators in the conduction band and the $j$th
valance band. $\sigma$ denotes the polarization components of the
light, and the Rabi frequency $\Omega_{\sigma,j}$ is proportional to
the laser amplitude and the interband optical matrix element.
Eq.(\ref{H_nocoulomb}) is diagonalized by a linear transformation,
$H_{\rm RF} =\sum_{{\bf k},m} \tilde{\varepsilon}_{m}({\bf k})
\tilde{a}^{\dagger}_{m,{\bf k}} \tilde{a}_{m,{\bf k}}$, where
$\tilde{a}_{m,{\bf k}}$ denotes the creation of a particle in the
photon-dressed band with the new dispersion
$\tilde{\varepsilon}_{m}({\bf k})$. We still use the subscript $c$
to denote the dressed band with highest energy in the RF,
$\tilde{a}_{c,{\bf k}} \equiv \alpha({\bf k}) a_{c,{\bf k}} + \sum_j
\beta_j({\bf k}) a_{{v_j},{\bf k}} $.

In absence of the laser field, the n-doped system has a filled
$k$-space region $A$ in the conduction band. It is convenient to use
a hole picture in which the above configuration reads $| G \rangle =
\prod_{{\bf k} \notin A} b^{\dagger}_{c,-{\bf k}} |\Phi_0 \rangle $.
$|\Phi_0 \rangle$ denotes complete filling of all bands, and
$b^{\dagger}_{c,-{\bf k}} \equiv a_{c,{\bf k}}$ is hole creation
operator for unfilled states. Without conduction band spin-orbit
coupling, the states $b^{\dagger}_{c,-{\bf k}} |\Phi_0 \rangle$ have
zero Berry curvatures and hence $|G\rangle$ has no topological
contribution to the Hall conductance. If the laser field,
red-detuned from the absorption edge $\omega_{\rm edge} \equiv {\rm
min}_{{\bf k} \notin A} (\varepsilon_c({\bf k}) -
\varepsilon_{v_j}({\bf k}))$, is adiabatically switched on, the
system will evolve from $|G \rangle$ to the dressed state $|
\tilde{G} \rangle = \prod_{{\bf k} \notin \tilde{A}}
\tilde{b}^{\dagger}_{c,{-\bf k}} |\Phi_0 \rangle $ where
$\tilde{b}^{\dagger}_{c,-{\bf k}} \equiv \tilde{a}_{c,{\bf k}}$. We
have assumed elastic scattering by impurities and temperatures low
enough ($< 100$ K) so that the phonon bath cannot supply the energy
to excite the electron out of the lower dressed bands. Weak
spontaneous emission of photons by the virtual excitations is also
neglected for the moment. In the cases considered, the new
equilibrium $k$-space filling $\tilde{A}$ (the lowest energy
configuration in the RF) differs only slightly from $A$. Because of
the admixture of the valance bands which have strong spin-orbit
coupling, the dressed band $\tilde{b}^{\dagger}_{c,{\bf k}} |\Phi_0
\rangle $ has non-zero Berry curvature ${\bf F}^c({\bf k}) =
\boldsymbol{\nabla}_{\bf k} \times {\bf R}^c({\bf k})$ where ${\bf
R}^c \equiv i\langle \Phi_0 | \tilde{b}_{c,{\bf k}} e^{i \bf{k}
\cdot \bf{r}} (\partial/\partial {\bf k})  e^{-i \bf{k} \cdot
\bf{r}} \tilde{b}^{\dagger}_{c,{\bf k}} |\Phi_0 \rangle
$~\cite{Chang_BerryPhase}. The dressed state $| \tilde{G} \rangle$
thus acquires topological transport properties.

In a direct bandgap III-V semiconductor, the electronic states
affected by light are those near the $\Gamma$ point. It is a good
approximation to neglect interband scattering and interband exchange
terms in Coulomb interaction since the periodic part of Bloch
functions are orthogonal between different bands near the $\Gamma$
point~\cite{OpticalStark}. For the same reason, we also assume no
interband scattering by phonons and smooth disorder. Thus, in the
RF, the Coulomb interactions and the coupling to phonon and impurity
potentials has the time-independent form.

Coulomb interactions which have several major effects are well
accounted for in optically excited semiconductors by the Hartree
Fock approximation~\cite{OpticalStark}. First, the band dispersion
$\varepsilon_l({\bf k})$ includes the interaction energies between
the occupied states in the ground state configuration $| G \rangle$.
Second, considering the interaction between the virtual
excitations~\cite{OpticalStark}, the transitions energies are
further renormalized, $\varepsilon_c({\bf k}) \rightarrow
\varepsilon_c({\bf k}) - \sum_{j,{\bf k^{\prime}} \notin \tilde{A}}
V({\bf k}^{\prime}- {\bf k}) |\beta_j({\bf k}^{\prime})|^2$ and
$\varepsilon_{v_j}({\bf k}) \rightarrow \varepsilon_{v_j}({\bf k}) +
\sum_{{\bf k^{\prime}} \notin \tilde{A}} V({\bf k}^{\prime} - {\bf
k}) |\beta_j({\bf k}^{\prime})|^2$, where $V({\bf k}^{\prime} - {\bf
k})$ is the screened Coulomb interaction. The effective Rabi
frequency is also renormalized $\Omega_{\sigma,v_j}({\bf k})
\rightarrow \Omega_{\sigma,v_j}({\bf k}) + \sum_{{\bf k^{\prime}}
\notin \tilde{A}} V({\bf k}^{\prime} - {\bf k}) \alpha ({\bf
k}^{\prime}) \beta_j^{\ast}({\bf k}^{\prime})$. The diagonalization
of Eq.(\ref{H_nocoulomb}) has to be performed self-consistently. We
also note that the optical absorption edge becomes $\omega_{\rm
edge} \equiv {\rm min}_{{\bf k} \notin A} (\varepsilon_c({\bf k}) -
\varepsilon_{v_j}({\bf k}) -\epsilon_b)$ where $\epsilon_b$ is the
binding energy to form an exciton.

\begin{figure}[t]
\includegraphics[width=7cm, height=5.4cm,bb=0 0 517 417]{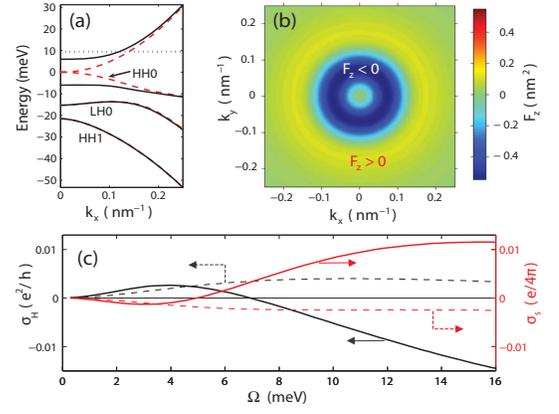}
\caption{(Color online) (a) Dispersion relations of the dressed
bands (solid curves) in the $\mathcal{H}_-$ subspace with a
$\sigma+$ polarized laser ($\Omega \equiv |\Omega_{+,\rm{HH0}}|=6$
meV). Dashed curves denotes the undressed bands in the rotating
frame. The horizontal dotted line denote the Fermi energy. (b) The
corresponding Berry Curvature distribution of the top dressed band
$\tilde{b}^{\dagger}_{c,{\bf k}} |\Phi_0 \rangle $. (c) Charge Hall
conductance $\sigma_{\rm H}$ (black curves) and pseudo-spin Hall
conductance $\sigma_s$ (red curves) as a function of laser strength
$\Omega$. Dashed curves denote the clean contribution and solid
curves include disorder effects. $k_B T =0.5$ meV, $\omega_0=E_g$,
and the electron density is fixed at $n=9\times 10^{12}
~\text{cm}^{-2}$.} \label{fig1}
\end{figure}

We consider an n-doped GaAs/AlGaAs QW, with symmetric confinement in
the growth ($z$) direction. For the conduction band electrons,
Rashba spin-orbit coupling vanishes since there is no SIA, and we
neglect the weak Dresselhaus spin-orbit coupling except for the
consequence of a finite spin relaxation time in the presence of
phonon scattering~\cite{SpinT1_QW}. QW confinement splits the
$\Gamma_6$ ($S=1/2$) bulk conduction band into subbands denoted as
$|S_z, {\bf k}_{\perp}, l_c \rangle$ where $S_z=\pm1/2$ is the spin
$z$ quantum number, ${\bf k}_{\perp}$ the in-plane wavevector, and
$l$ the quantum number of the QW confinement. The $\Gamma_8$
($J=3/2$) valance bands in bulk material is described by the
standard Luttinger Hamiltonian~\cite{Broido_Sham_holemass}. The
eigenstates of $\hat{J}_z$ together with the QW confinement define a
basis sets: $|J_z, {\bf k}_{\perp}, l_v \rangle$, similar to that of
the conduction subbands. Off-diagonal terms in the Luttinger
Hamiltonian are the origin of $k$-space Berry curvature, and they
couple the valance subbands at the same $\bf{k}_{\perp}$ with the
selection rule: (i) $\Delta J_z=\pm 1$ and $\Delta l_v$ odd (a
change of parity in $z$ direction); or (ii) $\Delta J_z=\pm 2$ and
$\Delta l_v$ even (no parity change). A light field with $\sigma$
polarization propagating in the $z$-direction couples the conduction
and valance subbands with the selection rules: $ S_z= J_z + \sigma$
and $l_c-l_v$ even. Therefore, we always have two decoupled
subspaces: (i) $\mathcal{H}_+ = \{ |S_z=1/2, l_c= e \rangle,
|S_z=-1/2, l_c= o \rangle, |J_z=3/2, l_v= e \rangle, |J_z=-1/2, l_v=
e \rangle, |J_z=-3/2, l_v=o \rangle, |J_z=1/2, l_v=o \rangle \} $
where we have used $e$ ($o$) to denote even (odd) integer; (ii)
$\mathcal{H}_-$ is the time reversal of $\mathcal{H}_+$. For typical
QW, the inclusion of lowest conduction subbands with $l_c=0$,
light-hole subbands with $l_v=0$, and heavy-hole subbands with
$l_v=0,1$ suffices since all other subbands are faraway in energy.

\begin{figure}[t]
\includegraphics[width=7cm, height=5.1cm,bb=0 0 517 389]{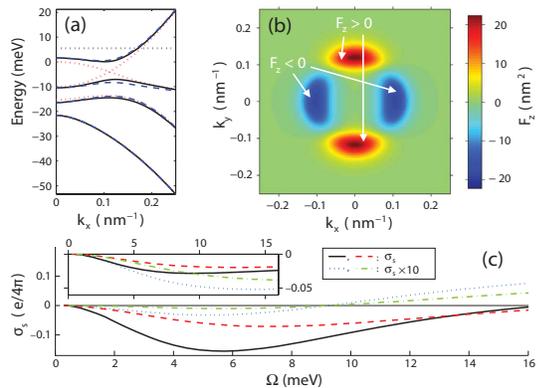}
\caption{(Color online) (a) Dispersion relations of the dressed
bands in subspace $\mathcal{H}_-$ with a $X$ polarized laser
($\omega_0-E_g=10$ meV and $\Omega = 6$ meV). Solid (dashed) curves
denote the band dispersions along $k_x=0$ ($k_y=0$). The Red dotted
curves denotes the undressed bands in the RF. The horizontal dotted
line denote the Fermi energy ($n=18\times 10^{12}~\text{cm}^{-2}$).
(b) The corresponding Berry Curvature distribution of the top
dressed band $\tilde{b}^{\dagger}_{c,{\bf k}} |\Phi_0 \rangle $. (c)
Pseudo-spin Hall conductance as a function of laser Rabi frequency
$\Omega$. The inset shows the clean contribution. The parameters
are, respectively, {\it solid curve}: $\omega_0-E_g=13$ meV and
$n=18\times 10^{12}~\text{cm}^{-2}$; {\it dashed}: $\omega_0-E_g=10$
meV and $n=18\times 10^{12}~\text{cm}^{-2}$; {\it dotted}:
$\omega_0=E_g$ and $n=9\times 10^{12}~\text{cm}^{-2}$; {\it
dash-dotted}: $\omega_0-E_g=-4$ meV, and $n=9\times
10^{12}~\text{cm}^{-2}$. The latter two curves are horizontally
scaled by a factor of $10$. $k_B T =0.5$ meV.} \label{fig2}
\end{figure}

Numerical evaluations were performed for a QW with thickness $d=10$
nm. For the QW layer and the barrier, we assume the same Luttinger
parameters ($\gamma_1=6.85$, $\gamma_2=2.1$, and $\gamma_3=2.9$) and
conduction band effective mass ($m_c=0.078 m_e$). We simply assume
$x$, $y$, and $z$ axes are the [100], [010] and [001] directions.
The cubic anisotropy gives a weak angular dependence on the scale of
several percent. The valance barrier height has been taken equal to
$140$ meV. The electron gas remains degenerate over the temperature
range we considered ($< 100$ K). The Rabi frequencies $\Omega_{\mp,
{\rm HH0}}$ for the transition between $|J_z= \pm 3/2, l_v= 0
\rangle$ and $|S_z=\pm1/2, l_c= 0 \rangle$ subbands are used to
index the laser strength, $\Omega \equiv
(|\Omega_{+,\rm{HH0}}|^2+|\Omega_{-,\rm{HH0}}|^2)^{1/2}$. For the
Coulomb interaction, we follow the treatment
of~\cite{2DCoulomb_Eisenstein_simplified}.

Fig.\ref{fig1}(a) shows the band dispersions with and without
dressing by $\sigma+$ polarized light with frequency tuned in
resonance with the bandgap $E_g \equiv \varepsilon_c(0) -
\varepsilon_{\rm HH0}(0)$ (note that $\omega_{\rm edge}$ is $\sim
10$ meV above $E_g$). The states near the $\Gamma$ point are
strongly affected and the top dressed band acquires $k$-space Berry
curvature with an $s$-like distribution [Fig.\ref{fig1}(b)]. The
negative Berry curvature distribution for small $k$ is compensated
by positive curvature for large $k$, so the Chern number of the top
dressed band is still $0$. Nonetheless, there is a nonzero {\it
clean} contribution to the carrier Hall conductance in each subspace
$\mathcal{H}_{\pm}$ : $\sigma^{\rm cl}_{\pm} = e^2/\hbar \sum_{\bf
k} (1-f) F^c_z({\bf k})$, where $f$ gives the Fermi distribution of
particles in the top dressed band. Since circularly polarized light
breaks time reversal symmetry, there is a net charge Hall
conductance $\sigma_{\rm H} \equiv \sigma_{+} + \sigma_{-}$ with the
magnitude controlled by the strength and frequency detuning of the
laser [Fig.~\ref{fig1} (c)].

In a realistic crystal, we need to consider disorder effects. In a
heavily modulation doped QW, the dominant cause of disorder is from
dopants embedded in the barrier layers with smooth potentials in the
QW layer. Electrons acquire an anomalous coordinate shift with
magnitude proportional to the Berry curvature $F^c_z({\bf k})$ when
they scatter off a smooth impurity potential~\cite{Sidejump_smooth}.
This leads to a side-jump contribution to the Hall conductance,
$\sigma_{\pm}^{\rm sj}= - e^2/\hbar \sum_{{\bf k}} F^c_z({\bf k})
\frac{\partial f}{\partial \tilde{\varepsilon}_c} \frac{\partial
\tilde{\varepsilon}_c}{\partial {\bf k}} \cdot {\bf k} $,
independent of the scattering rate~\cite{Berger_Sidejump}. The
admixture of valance band component also causes skew-scattering of
conduction electrons by impurities~\cite{Smit_skew1}, but its
contribution to the Hall conductance is inversely proportional to
the scattering rate and therefore is suppressed in this heavily
doped case~\cite{Weitering}. The total Hall conductance is then
$\sigma^{\rm tot} \equiv \sigma^{\rm cl} + \sigma^{\rm sj}$. Since
$F^c_z({\bf k})$ has strong $k$-dependence which changes with the
light intensity, $\sigma^{\rm sj}$ has a more drastic change as a
function of $\Omega$ and $\sigma^{\rm tot}$ can even change sign
[Fig.\ref{fig1} (c)].

With linearly polarized light, the $k$-space Berry curvature in the
top dressed band has a $p$-like distribution [Fig.\ref{fig2} (b)].
The carrier Hall conductances in the two subspaces
$\mathcal{H}_{\pm}$ have equal magnitude but opposite sign,
$\sigma_+ = -\sigma_-$, so the net charge Hall conductance vanishes.
We define a pseudo-spin Hall conductance
$\sigma_s=\hbar/(2e)(\sigma_{+} - \sigma_{-})$ and its dependence on
the laser strength and frequency detuning is studied in
Fig.\ref{fig2} (c). We note that with circular polarized light, the
charge Hall current typically has a majority contribution from one
of the subspaces, so that there is a finite pseudo-spin Hall
conductance as well [Fig.\ref{fig1}(c)].

\begin{figure}[t]
\includegraphics[width=7.5cm, height=4.4cm,bb=39 298 582 650]{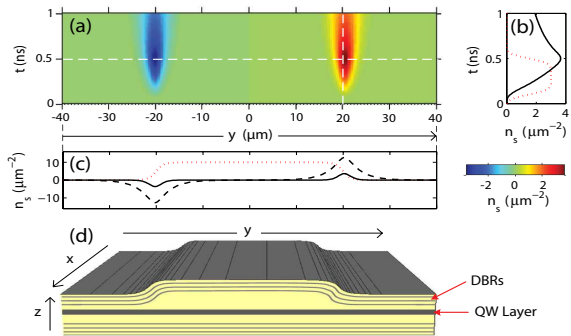}
\caption{(Color online) (a) Spatial and time dependence of spin
accumulation $n_s$ driven by a DC electric field in the
$x$-direction ($1~\text{mV}/\mu \text{m}$). The SHE is activated by
a X-polarized laser pulse with central frequency $\omega_0-E_g=13$
meV. The temporal and spatial profile of the pulse is shown by the
dotted curves in (b) and (c), with the peak strength at $y=0$ being
$\Omega=5$ meV. The system is assumed to be translationally
invariant in $x$ direction in the region of interest. (b) Time
dependence of the spin density (solid curve) at $y=20~\mu$m. (c)
Solid curve denotes spatial dependence of spin density at $t=0.5$
ns. The dashed curve denotes the steady state spin accumulation
driven by a CW laser with the same spatial profile and
$\Omega(y=0)=5$ meV. (d) Schematic illustration of a QW embedded in
an optical cavity~\cite{DotinCavity_Shih}. The lowest cavity mode
has spatial profile shown by the dotted curve in (c).} \label{fig3}
\end{figure}

The sharpness of the spatial variation of the laser intensity is
bounded by the optical wavelength $\lambda \sim \mu$m which is
smooth enough to guarantee adiabatic evolution between the dressed
$\tilde{b}^{\dagger}_{c,{\bf k}} |\Phi_0 \rangle$ and undressed
$b^{\dagger}_{c,{\bf k}} |\Phi_0 \rangle$ at the light irradiation
edges. The difference in the carrier Hall conductance $\sigma_{\pm}$
inside and outside the irradiation area leads to carrier
accumulations $\Delta n_{\pm}$ at the edges in each subspace
$\mathcal{H}_{\pm}$, determined by the drift-diffusion equations:
$\frac{\partial }{\partial t}\Delta n_{\pm }-\frac{1}{e}
\frac{\partial }{\partial y } \sigma_{\pm} E_x- D \frac{\partial
^{2}}{\partial y^{2}} \Delta n_{\pm} =-\frac{ \Delta n_{\pm } -
\Delta n_{\mp }}{\tau _{s}}$, where $\tau_s$ is the spin relaxation
time, $L_s\equiv \sqrt{D \tau_s}$ the spin diffusion length, and
momentum relaxation is assumed orders faster than $\tau_s$. For a
sufficiently large electron density $n \gg \Delta n_{\pm}$, the
carrier accumulations are all in the top dressed bands. We define
the pseudo-spin density $n_s=\Delta n_+ - \Delta
n_-$~\cite{SHE_Awschalom}. Outside the irradiation area or after the
light is adiabatically switched off, $n_s$ just gives the pure spin
accumulation of the conduction band electrons. In Fig.\ref{fig3}
(a-c), we show $n_s$ generated by a laser pulse with duration $\sim
0.5$ ns and a CW laser, assuming $\tau_s = 5$ ns~\cite{SpinT1_QW}
and $L_s = 4~\mu$m~\cite{SHE_Awschalom}.

In summary, we have shown that a sizable AHE and SHE due to dressed
band topology can be optically activated in n-doped paramagnetic
semiconductors. The absence of spin-orbit interactions when the
light is switched off enables long spin lifetimes. With a reasonable
value of the laser Rabi frequency ($\sim $ meV), optically activated
topological spin Hall transport can be 3 orders stronger than that
from skew scattering~\cite{SHE_Awschalom}. The laser excitation is
operated in the reactive regime where the system excitation is
predominantly virtual. Spontaneous emission of a photon at frequency
$\omega_0 - \tilde{\varepsilon}_c({\bf k}) +
\tilde{\varepsilon}_{v}({\bf k})$ (typically $> 10$ meV below the
laser resonance) can bring $\tilde{b}^{\dagger}_{c,{-\bf k}} |\Phi_0
\rangle$ to $\tilde{b}^{\dagger}_{v,{-\bf k}} |\Phi_0
\rangle$~\cite{OpticalStark}. Thus, in free space, the system can
have weak but finite absorption of energy from the light. For laser
pulses with duration shorter than the spontaneous emission time
($\sim $~ns), the energy absorption would be
small~\cite{OpticalStark}. Embedding the QW in an optical cavity
with a moderate $Q\sim 100$ [see Fig.~\ref{fig3}(d)] can eliminate
this unwanted dissipation. The cavity photon modes are lower bounded
in frequency, and the spectrum edge, by design, is at the frequency
of interest which allows in-couple of the laser and forbids
out-couple at the spontaneous emission frequency. Cavity enhancement
also allows a large Rabi frequency $\Omega$ to be achieved with a
relatively weak external laser field.

We acknowledge useful discussions with X. F. Song, D. Xiao, X. Li
and C. K. Shih. The work was supported by NSF, DOE and the Welch
Foundation.


\begin{thebibliography}{30}
\expandafter\ifx\csname
natexlab\endcsname\relax\def\natexlab#1{#1}\fi
\expandafter\ifx\csname bibnamefont\endcsname\relax
  \def\bibnamefont#1{#1}\fi
\expandafter\ifx\csname bibfnamefont\endcsname\relax
  \def\bibfnamefont#1{#1}\fi
\expandafter\ifx\csname citenamefont\endcsname\relax
  \def\citenamefont#1{#1}\fi
\expandafter\ifx\csname url\endcsname\relax
  \def\url#1{\texttt{#1}}\fi
\expandafter\ifx\csname urlprefix\endcsname\relax\def\urlprefix{URL
}\fi \providecommand{\bibinfo}[2]{#2}
\providecommand{\eprint}[2][]{\url{#2}}

\bibitem{ORKKY}
\bibinfo{author}{\bibfnamefont{C.}~\bibnamefont{Piermarocchi}}~{\it et al.},
  \bibinfo{journal}{Phys.\ Rev. Lett.} \textbf{\bibinfo{volume}{89}},
  \bibinfo{pages}{167402} (\bibinfo{year}{2002});
\bibinfo{author}{\bibfnamefont{J.}~\bibnamefont{Fernandez-Rossier}}~{\it et al.},
  {\it ibid.} \textbf{\bibinfo{volume}{93}},
  \bibinfo{eid}{127201} (\bibinfo{year}{2004}).

\bibitem[{\citenamefont{Ganichev et~al.}(2002)\citenamefont{Ganichev, Ivchenko,
  Bel'kov, Tarasenko, Sollinger, Weiss, Wegscheider, and
  Prettl}}]{Spingalvanic_Ganichev}
\bibinfo{author}{\bibfnamefont{S.~D.} \bibnamefont{Ganichev}}~{\it et al.},
  \bibinfo{journal}{Nature} \textbf{\bibinfo{volume}{417}},
  \bibinfo{pages}{153} (\bibinfo{year}{2002}).

\bibitem[{\citenamefont{Stevens et~al.}(2003)\citenamefont{Stevens, Smirl,
  Bhat, Najmaie, Sipe, and van Driel}}]{Photonspincurrent_Smirl}
\bibinfo{author}{\bibfnamefont{M.~J.} \bibnamefont{Stevens}}~{\it et al.},
  \bibinfo{journal}{Phys. Rev. Lett.} \textbf{\bibinfo{volume}{90}},
  \bibinfo{eid}{136603} (\bibinfo{year}{2003}).

\bibitem{Photoinduced_Hall_Zhang}
\bibinfo{author}{\bibfnamefont{J.}~\bibnamefont{Li}}~{\it et al.},
  \bibinfo{journal}{Appl. Phys. Lett.} \textbf{\bibinfo{volume}{88}},
  \bibinfo{pages}{162105} (\bibinfo{year}{2006});
\bibinfo{author}{\bibfnamefont{X.}~\bibnamefont{Dai}} \bibnamefont{and}
\bibinfo{author}{\bibfnamefont{F.~C.} \bibnamefont{Zhang}},
  \bibinfo{journal}{cond-mat/0605159}  (\bibinfo{year}{2006}).

\bibitem[{\citenamefont{Comte and Mahler}(1986)}]{OpticalStark}
\bibinfo{author}{\bibfnamefont{C.}~\bibnamefont{Comte}} \bibnamefont{and}
  \bibinfo{author}{\bibfnamefont{G.}~\bibnamefont{Mahler}},
  \bibinfo{journal}{Phys. Rev. B} \textbf{\bibinfo{volume}{34}},
  \bibinfo{pages}{7164} (\bibinfo{year}{1986});
\bibinfo{author}{\bibfnamefont{S.}~\bibnamefont{Schmitt-Rink}},
  \bibinfo{author}{\bibfnamefont{D.~S.} \bibnamefont{Chemla}},
  \bibnamefont{and} \bibinfo{author}{\bibfnamefont{H.}~\bibnamefont{Haug}},
  {\it ibid.} \textbf{\bibinfo{volume}{37}},
  \bibinfo{pages}{941} (\bibinfo{year}{1988});
\bibinfo{author}{\bibfnamefont{S.}~\bibnamefont{Glutsch}} \bibnamefont{and}
  \bibinfo{author}{\bibfnamefont{R.}~\bibnamefont{Zimmermann}},
  {\it ibid.} \textbf{\bibinfo{volume}{45}},
  \bibinfo{pages}{5857} (\bibinfo{year}{1992}).

\bibitem{QHE_TKNN}
\bibinfo{author}{\bibfnamefont{D.~J.} \bibnamefont{Thouless}}~{\it et al.},
  \bibinfo{journal}{Phys. Rev. Lett.} \textbf{\bibinfo{volume}{49}},
  \bibinfo{pages}{405} (\bibinfo{year}{1982}).

\bibitem[{\citenamefont{Chang and Niu}(1996)}]{Chang_BerryPhase}
\bibinfo{author}{\bibfnamefont{M.-C.} \bibnamefont{Chang}} \bibnamefont{and}
  \bibinfo{author}{\bibfnamefont{Q.}~\bibnamefont{Niu}},
  \bibinfo{journal}{Phys. Rev. B} \textbf{\bibinfo{volume}{53}},
  \bibinfo{pages}{7010} (\bibinfo{year}{1996}).

\bibitem[{\citenamefont{Karplus and Luttinger}(1954)}]{Karplus_AHE}
\bibinfo{author}{\bibfnamefont{R.}~\bibnamefont{Karplus}} \bibnamefont{and}
  \bibinfo{author}{\bibfnamefont{J.~M.} \bibnamefont{Luttinger}},
  \bibinfo{journal}{Phys. Rev.} \textbf{\bibinfo{volume}{95}},
  \bibinfo{pages}{1154} (\bibinfo{year}{1954}).

\bibitem[{\citenamefont{Smit}(1955)}]{Smit_skew1}
\bibinfo{author}{\bibfnamefont{J.}~\bibnamefont{Smit}},
  \bibinfo{journal}{Physica} \textbf{\bibinfo{volume}{21}},
  \bibinfo{pages}{877} (\bibinfo{year}{1955});
  {\it ibid.} \textbf{\bibinfo{volume}{24}}, \bibinfo{pages}{39}
  (\bibinfo{year}{1958}).

\bibitem[{\citenamefont{Berger}(1970)}]{Berger_Sidejump}
\bibinfo{author}{\bibfnamefont{L.}~\bibnamefont{Berger}},
  \bibinfo{journal}{Phys. Rev. B} \textbf{\bibinfo{volume}{2}},
  \bibinfo{pages}{4559} (\bibinfo{year}{1970}).

\bibitem[{\citenamefont{Jungwirth et~al.}(2002)\citenamefont{Jungwirth, Niu,
  and MacDonald}}]{AHE_FMSM_Jungwirth}
\bibinfo{author}{\bibfnamefont{T.}~\bibnamefont{Jungwirth}},
  \bibinfo{author}{\bibfnamefont{Q.}~\bibnamefont{Niu}}, \bibnamefont{and}
  \bibinfo{author}{\bibfnamefont{A.~H.} \bibnamefont{MacDonald}},
  \bibinfo{journal}{Phys. Rev. Lett.} \textbf{\bibinfo{volume}{88}},
  \bibinfo{pages}{207208} (\bibinfo{year}{2002}).

\bibitem[{\citenamefont{Fang et~al.}(2003)\citenamefont{Fang, Nagaosa,
  Takahashi, Asamitsu, Mathieu, Ogasawara, Yamada, Kawasaki, Tokura, and
  Terakura}}]{AHE_Fang}
\bibinfo{author}{\bibfnamefont{Z.}~\bibnamefont{Fang}}~{\it et al.},
  \bibinfo{journal}{Science} \textbf{\bibinfo{volume}{302}},
  \bibinfo{pages}{92} (\bibinfo{year}{2003}).

\bibitem[{\citenamefont{Yao et~al.}(2004)\citenamefont{Yao, Kleinman,
  MacDonald, Sinova, Jungwirth, Wang, Wang, and Niu}}]{AHE_Niu}
\bibinfo{author}{\bibfnamefont{Y.}~\bibnamefont{Yao}}~{\it et al.},
  \bibinfo{journal}{Phys. Rev. Lett.} \textbf{\bibinfo{volume}{92}},
  \bibinfo{pages}{037204} (\bibinfo{year}{2004}).

\bibitem[{\citenamefont{Onoda et~al.}(2006)\citenamefont{Onoda, Sugimoto, and
  Nagaosa}}]{AHE_ins_ext_Nagaosa}
\bibinfo{author}{\bibfnamefont{S.}~\bibnamefont{Onoda}},
  \bibinfo{author}{\bibfnamefont{N.}~\bibnamefont{Sugimoto}}, \bibnamefont{and}
  \bibinfo{author}{\bibfnamefont{N.}~\bibnamefont{Nagaosa}},
  \bibinfo{journal}{Phys. Rev. Lett.} \textbf{\bibinfo{volume}{97}},
  \bibinfo{pages}{126602} (\bibinfo{year}{2006}).

\bibitem[{\citenamefont{Murakami et~al.}(2003)\citenamefont{Murakami, Nagaosa,
  and Zhang}}]{SHE_Murakami}
\bibinfo{author}{\bibfnamefont{S.}~\bibnamefont{Murakami}},
  \bibinfo{author}{\bibfnamefont{N.}~\bibnamefont{Nagaosa}}, \bibnamefont{and}
  \bibinfo{author}{\bibfnamefont{S.-C.} \bibnamefont{Zhang}},
  \bibinfo{journal}{Science} \textbf{\bibinfo{volume}{301}},
  \bibinfo{pages}{1348} (\bibinfo{year}{2003});
\bibinfo{author}{\bibfnamefont{J.}~\bibnamefont{Sinova}}~{\it et al.},
\bibinfo{journal}{Phys. Rev. Lett.}
  \textbf{\bibinfo{volume}{92}}, \bibinfo{pages}{126603}
  (\bibinfo{year}{2004}).

\bibitem{SHE_Awschalom}
\bibinfo{author}{\bibfnamefont{Y.~K.} \bibnamefont{Kato}}~{\it et al.},
    \bibinfo{journal}{Science}
  \textbf{\bibinfo{volume}{306}}, \bibinfo{pages}{1910} (\bibinfo{year}{2004}).

\bibitem{SHE_Wunderlich}
\bibinfo{author}{\bibfnamefont{J.}~\bibnamefont{Wunderlich}}~{\it et al.},
  \bibinfo{journal}{Phys. Rev. Lett.} \textbf{\bibinfo{volume}{94}},
  \bibinfo{pages}{047204} (\bibinfo{year}{2005}).

\bibitem{Spincurrent_Niu}
\bibinfo{author}{\bibfnamefont{J.}~\bibnamefont{Shi}}~{\it et al.},
  \bibinfo{journal}{Phys. Rev. Lett.} \textbf{\bibinfo{volume}{96}},
  \bibinfo{pages}{076604} (\bibinfo{year}{2006}).

\bibitem[{\citenamefont{Bychkov and Rashba}(1984)}]{Rashba_1984}
\bibinfo{author}{\bibfnamefont{Y.~A.} \bibnamefont{Bychkov}} \bibnamefont{and}
  \bibinfo{author}{\bibfnamefont{E.~I.} \bibnamefont{Rashba}},
  \bibinfo{journal}{J. Phys. C: Solid State Phys.}
  \textbf{\bibinfo{volume}{17}}, \bibinfo{pages}{6039} (\bibinfo{year}{1984}).

\bibitem{twospace}
  \bibinfo{author}{\bibfnamefont{B.~A.}~\bibnamefont{Bernevig}},
  \bibinfo{author}{\bibfnamefont{T.~L.}~\bibnamefont{Hughes}}, \bibnamefont{and}
  \bibinfo{author}{\bibfnamefont{S.-C.} \bibnamefont{Zhang}},
  \bibinfo{journal}{Science} \textbf{\bibinfo{volume}{314}},
  \bibinfo{pages}{1757} (\bibinfo{year}{2006});
\bibinfo{author}{\bibfnamefont{E.~M.} \bibnamefont{Hankiewicz}},
  \bibinfo{author}{\bibfnamefont{G.}~\bibnamefont{Vignale}}, \bibnamefont{and}
  \bibinfo{author}{\bibfnamefont{M.~E.} \bibnamefont{Flatt\'e}},
  \bibinfo{journal}{Phys. Rev. Lett.} \textbf{\bibinfo{volume}{97}},
  \bibinfo{pages}{266601} (\bibinfo{year}{2006}).



\bibitem[{\citenamefont{Ohno et~al.}(1999)\citenamefont{Ohno, Terauchi, Adachi,
  Matsukura, and Ohno}}]{SpinT1_QW}
\bibinfo{author}{\bibfnamefont{Y.}~\bibnamefont{Ohno}}~{\it et al.},
  \bibinfo{journal}{Phys. Rev. Lett.} \textbf{\bibinfo{volume}{83}},
  \bibinfo{pages}{4196} (\bibinfo{year}{1999}).

\bibitem[{\citenamefont{Broido and Sham}(1985)}]{Broido_Sham_holemass}
\bibinfo{author}{\bibfnamefont{D.~A.} \bibnamefont{Broido}} \bibnamefont{and}
  \bibinfo{author}{\bibfnamefont{L.~J.} \bibnamefont{Sham}},
  \bibinfo{journal}{Phys. Rev. B} \textbf{\bibinfo{volume}{31}},
  \bibinfo{pages}{888} (\bibinfo{year}{1985}).

\bibitem[{\citenamefont{Eisenstein}(2005)}]{2DCoulomb_Eisenstein_simplified}
\bibinfo{author}{\bibfnamefont{J.~P.} \bibnamefont{Eisenstein}}, in
  \emph{\bibinfo{booktitle}{Nanophysics: Coherence and Transport}},
  edited by H. Bouchiat~{\it et al.}
  (\bibinfo{publisher}{Elsevier Science}, \bibinfo{year}{2005}).

\bibitem{Sidejump_smooth}
\bibinfo{author}{\bibfnamefont{N.~A.} \bibnamefont{Sinitsyn}}~{\it et al.},
  \bibinfo{journal}{Phys. Rev. B} \textbf{\bibinfo{volume}{72}},
  \bibinfo{eid}{045346} (\bibinfo{year}{2005}).

\bibitem{Weitering} Skew-scattering contribution can also be seperated
from intrinsic and side-jump contributions by their different
correlation with the longitudinal resistivity. See,
\bibinfo{author}{\bibfnamefont{C.} \bibnamefont{Zeng}}~{\it et al.},
  \bibinfo{journal}{Phys. Rev. Lett.} \textbf{\bibinfo{volume}{96}},
  \bibinfo{eid}{037204} (\bibinfo{year}{2006}).


\bibitem[{\citenamefont{Muller et~al.}(2006)\citenamefont{Muller, Shih, Ahn,
  Lu, and Deppe}}]{DotinCavity_Shih}
\bibinfo{author}{\bibfnamefont{A.}~\bibnamefont{Muller}}~{\it et al.},
  \bibinfo{journal}{Opt. Lett.} \textbf{\bibinfo{volume}{31}},
  \bibinfo{pages}{528} (\bibinfo{year}{2006}).

\end{thebibliography}
\end{document}